\documentclass[epj,referee]{svjour}
\usepackage{latexsym}
\usepackage{graphics}
\usepackage{amsmath}
\begin{document}
  \title{Quantitative relations between corruption and economic factors}
  %%  \subtitle{Quantitative relations between corruption and economic factors}
  \author{Jia Shao$^{1}$, Plamen Ch. Ivanov$^{1,2}$, Boris Podobnik$^{1,3,4}$,
    H.~Eugene~Stanley$^{1}$
  }        % Do not remove
%
%\shortauthor{Jia Shao, {\it et al} }
%offprints{}          % Insert a name or remove this line
%
\institute{
$^1$ Center for Polymer Studies and Department of Physics,
  Boston University, Boston, MA 02215\\
$^2$ Institute of Solid State Physics, Bulgarian Academy of Sciences, 1784 Sofia, Bulgaria\\
$^3$ Zagreb School of Economics and Management, Zagreb, Croatia \\
$^4$ Faculty of Civil Engineering, University of Rijeka, Croatia 
}

\date{Received: February 27, 2007 }

%\date{ Printed: \today }

% {\bf  DRAFT}

\begin{abstract}
{\bf %  and key 
%parameters of economic performance, however, remains largely qualitative and not yet conclusive \cite{SA1,MAURO,WHEE,HINE,WEI,IMF,LEFF,HUNT}. 
We report quantitative relations 
between corruption level and economic factors, such as country wealth and foreign investment per capita, which are 
characterized by a power law spanning multiple scales of wealth and investments per capita. These relations hold for diverse countries, and also remain stable 
over different time periods. We also observe a negative correlation between 
level of corruption and long-term economic growth. We find similar results  
for two independent indices of corruption, suggesting that the relation between corruption and wealth does not depend on the specific measure of corruption. 
The functional relations we report have implications when assessing the relative level of corruption for two countries with comparable wealth, and for 
quantifying the impact of corruption on economic growth and foreign investments.
%The functional relations we report 
%here can have implications when determining the relative level of corruption for two countries with comparable wealth, and for 
%quantifying the impact of corruption when planning foreign investments and economic growth.
}
\end{abstract}
%\maketitle

 % end of PACS codes
 %end of abstact

\maketitle

\section{Introduction}
\label{intro}
Corruption influences important aspects of social and economic life. 
The level of corruption in a given country is widely believed to be an important factor to consider when projecting economic growth, estimating the 
effectiveness of the government administration, making decisions for strategic investments, and forming international policies. 
The relation 
between corruption level and key 
parameters of economic performance is largely qualitative \cite{SA1,MAURO,IMF,LEFF,HUNT,WHEE,HINE,WEI}. 
Corruption has become
increasingly important with the globalization of the international economic and political relations between countries, which has led various governmental 
and non-governmental organizations to search for adequate measures to quantify levels of corruption \cite{SA1,MAURO,KKM,KK,TREI,jain}. 
%The relation 
%between corruption level and key 
%parameters of economic performance is largely qualitative \cite{SA1,MAURO,WHEE,HINE,WEI,IMF,LEFF,HUNT}. 
%While corruption has always played a role in society, with the globalization of 
%the world economy there is increased interest in finding a relationship between level of corruption
%and characteristics of economic performance such as wealth, growth and international investment across
%different countries. 

Systematic studies of corruption have been hampered because of the complexity and secretive 
nature of corruption, making it difficult to quantify. There have been concerted 
efforts to introduce quantitative measures suitable for describing levels of corruption across diverse countries \cite{CI,CPI,CC}. 
However, a specific functional dependence between quantitative measures of corruption and economic performance has not been established. 

Previous studies have suggested 
a negative association between corruption level and country wealth \cite{SA1,MAURO,IMF}. There is active
debate concerning the relation between corruption level and economic growth \cite{BARD,lambs}. Some earlier studies suggest that 
corruption may help the most efficient firms bypass bureaucratic obstacles and rigid laws \cite{LEFF,HUNT} leading to a positive effect
on economic growth, while more recent works do not find a significant negative dependence 
between corruption and growth \cite{SA1,MAURO}. Further, studies of net flow of foreign investment report conflicting 
results. Some studies find no significant correlation between inward foreign investment and corruption level in host
countries \cite{WHEE,HINE}, while others indicate a negative association between corruption and foreign investments \cite{MAURO,WEI}. This debate reflects
the inherent complexity of the problem as countries in the world vary dramatically in their social and economic development \cite{SCHN}. Thus, an open question 
remains whether there is a general functional relation between corruption level and key aspects of the economic performance of different countries.

We develop and test the hypothesis that there may be a power-law dependence between corruption level and economic performance which holds across diverse countries
regardless of 
differences in specific country characteristics such as country wealth (defined in our paper as gross domestic product per capita) or 
foreign direct investment. Recent studies 
show that diverse social and economic systems exhibit scale invariant behavior --- e.g., size ranking and growth of firms, universities, urban centers,
 countries and even people's personal fortunes follow a power law over a broad range of scales \cite{MAKSE,AXTELL,MST,YKL,FDF,UNIVER,IVANOV,NEWMAN}. Since countries
in the world greatly
differ in their wealth and foreign investments, we test the possibility that there may be an 
underlying organization, such that the cross-country relations between corruption level and country wealth, and corruption level and foreign 
investments exhibit a significant negative correlation characterized by 
scale-invariant properties over multiple scales, and thus they can be described by power laws. Specifically, we test if 
this scale-invariant behavior remains stable over different time periods, as well as its validity for different subgroups of countries. Finally, we demonstrate 
a strong correlation between corruption level and past long-term economic growth.

\bigskip

\section{Data and Methods}
\label{empirical}

We analyze the Corruption Perceptions Index (CPI) \cite{CPI,method} introduced by {\it Transparency International} \cite{CPI}, a global 
civil organization supported by a wide network of government agencies, developmental organizations, foundations, public 
institutions, the private sector, and individuals. The CPI is a composite index based on independent surveys of business people and on
assessments of corruption in different countries provided 
by more than ten independent institutions around the world, including the {\it World Economic Forum}, {\it United Nations Economic Commission for Africa},
the {\it Economist Intelligence Unit}, the {\it International Institute for Management Development} \cite{method}. The CPI spans 10-year period 1996-2005.
The different surveys and assessments use diverse sampling frames and different methodologies. 
Some of the institutions consult a panel of experts to assess the level of corruption, while others, such as the {\it International Institute for 
Management Development} and the {\it Political and Economic Risk Consultancy}, turn to elite businessmen and businesswomen from different industries. 
Further, certain institutions gather information about the perceptions of corruption from {\it residents} with respect to the performance of their home
countries, while other institutions survey the perceptions of {\it non-residents} in regard to foreign countries or specifically in regard to 
neighboring countries.  All sources employ a homogeneous
definition of corruption as the misuse of public power for private benefit, such as bribing public
officials, kickbacks in public procurement, or embezzlement of public funds. Each of these sources also assesses the ``extent'' of corruption among
public officials and politicians in different countries. Transparency International uses non-parametric statistics for standardizing 
the data and for determining the precision of the scores \cite{method}.
While there is a certain subjectivity in people's perceptions of corruption, the large number of independent surveys and assessments based on different
methodologies averages out most of the bias. The CPI ranges from 0 (highly corrupt) to 10 (highly transparent).
%, is a composite index
%using data compiled from 16 surveys of businesspeople and assessments by country 
%analysts. It consists of credible sources using diverse
%sampling frames and different methodologies. All sources apply a common difinition of corruption such as the misuse of
%public power for private benefit, for exmple bribing of public officials, kickbacks in public
%procurement, or embezzlement of public funds. This criterion applies for both developing and 
%developed countries. Non-parametric statistics are used for standardizing the 
%the data and for determining the precision of the scores \cite{method}.
%The CPI
%ranges from 0 (highly corrupt) to 10 (highly transparent). 

We also analyze a different measure of corruption, the 
Control of Corruption Index (CCI) \cite{KKM,CC} provided by the {\it World Bank} \cite{CC}.
%The CCI is independent of the CPI, and thus allows us to test the validity of our results. 
The CCI ranges from --2.5 to 2.5, with positive numbers indicating low levels of corruption. As a measure of country wealth, we use 
 the {\it gdp}, defined to be the annual nominal gross domestic product per capita in 
current prices in U.S. dollars, provided by the {\it International 
Monetary Fund} (IMF) \cite{IMFGDP} over the 26-year period 1980-2005. As a measure of foreign direct investment we use annual
data from the {\it Bureau of Economic Analysis} \cite{BEA} of the United States (U.S.) government, which
represents the direct investment received by different countries 
from the U.S. over the period 2000-2004. These data are appropriate for our study since (i) the U.S. has been the dominant source of foreign investment
in the past decades and (ii) the 1977 Foreign Corrupt Practices Act (FCPA) \cite{FCPA} holds U.S. companies legally liable for bribing 
foreign government officials, which makes the U.S. a source country which penalizes its multinational companies for corruption practices \cite{WEI}.

\bigskip

\section{Results and Discussion}
\label{conclu}

\subsection{Relation Between Corruption Level and Country Wealth.} 
\begin{figure*}
  \centering
 \vspace{-1cm}
  \rotatebox{0}{\resizebox{1.1\columnwidth}{!}{\includegraphics{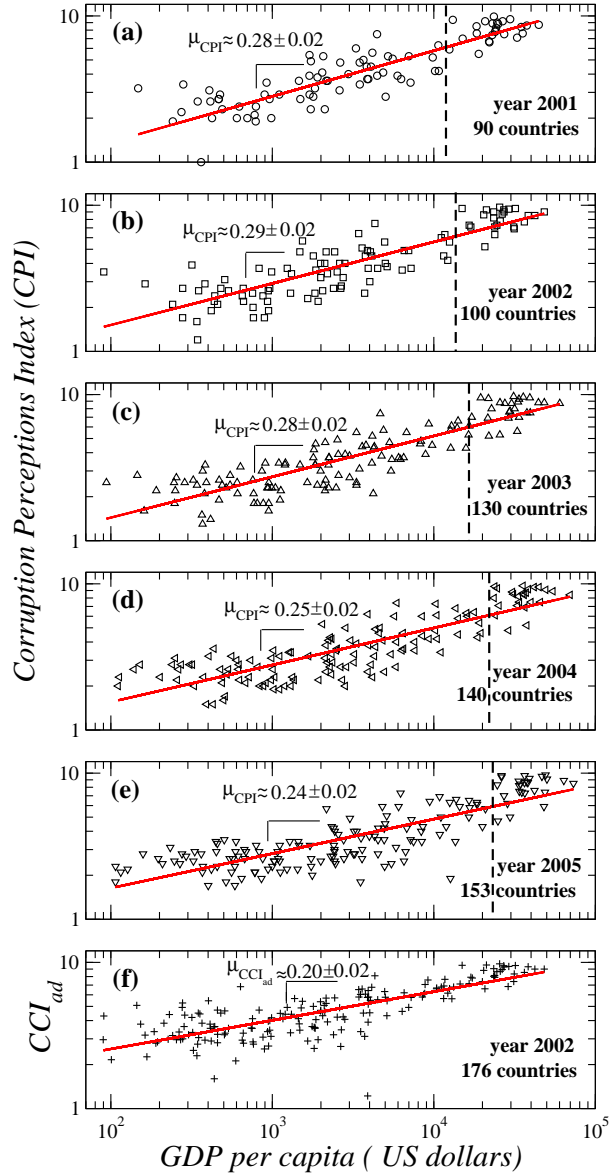}}}
 \vspace*{-1cm}
  \caption{ Log-log plots of the corruption perceptions index (CPI) versus GDP per capita ({\it gdp}) indicating a power-law functional 
            dependence. A low value of CPI corresponds to a high level of corruption \cite{CPI}. Data on 
            {\it gdp} are obtained as current prices in U.S. dollars \cite{IMFGDP}. 
            (a)-(e) The power-law functional
            dependence remains stable over different time periods,
            and is characterized by similar values of the exponent $\mu$ for different years and different number of countries.
            The power-law fit indicates the expected level of corruption expected for given country wealth. 
            Note that, comparing two countries with a similar {\it gdp}, the country placed above the power-law fit 
            is less corrupt than one would expect for its level of wealth, 
            while the country below the power-law curve has a relatively higher level of corruption than one would expect for its wealth.
            (f) We obtain similar results for the {\it adjusted} control of corruption index $CCI_{ad}$ \cite{CC}, which 
            is independent of CPI, indicating that the scale-invariant 
            relation between
            corruption and wealth does not depend on the specific measure of corruption. Vertical dashed lines in the panels seperate the top 30 wealthiest 
            countries.
            (see Fig. 5 and Fig. 6)}
            %Countries above the power-law fit are less corrupt relative to the average level of corruption expected 
            %for their wealth, while
            %countries below the power-law fit exhibit higher corruption levels than expected. 
            
    %(b) Log-log plot of CPI index versus  
    %GDP for 140 countries. 
    %(c) Log-log plot of CPI/(GDP) index versus  
    %GDP. 
    \label{fig.1}
\end{figure*}

To test if there is a common functional 
dependence between corruption level and country wealth, we plot the CPI versus {\it gdp} for different countries [Fig.~\ref{fig.1}(a-e)]. 
We find a positive correlation between CPI and country wealth, which can be well approximated by a power 
law 
\begin{equation}
      CPI\sim ({\it gdp})^{\mu},
\end{equation}
where $\mu>0$, indicating that richer 
countries are less corrupt. Most countries fall close to the power-law fitting line shown in Fig.~\ref{fig.1}, consistent with specific functional
relation between corruption and country wealth even for countries characterized by levels of wealth ranging over a factor of $10^{3}$. 
This finding in Eq. (1) indicates
 that the relative corruption level between two countries should be considered not only in terms of CPI values but also in the context of country wealth.
 For example, two countries with a large difference in their {\it gdp} on average will not have the same
level of corruption, as our results quantify the degree to which poorer countries with lower {\it gdp} have higher levels of corruption.

The quantitative relation between CPI and {\it gdp} for all countries in the world --- represented by the power-law fitting curves
in Fig. 1 --- indicates where is the ``expected'' level
of corruption for a given level of wealth. A country above (or below) the fitting line is less (or more) corrupt than expected for its level of wealth.
For example, comparing the relative corruption level of two countries with
similar {\it gdp} such as Bulgaria and Romania, one can assess
that Bulgaria is less corrupt than Romania [Fig. 3]. Depending whether a specific 
country is above (e.g., Bulgaria) or below (e.g., Romania) the power-law fit, one can assess if this 
country is less (or more) corrupt relative to the average level of corruption corresponding to the wealth of this country.

%\vspace*{-4cm}

Moreover, the quantitative dependence we find in Eq. (1) allows us to compare the relative levels of 
corruption between two countries which belong to two different wealth brackets.
Specifically, two countries with a very different {\it gdp} should not be compared only by the value of their CPI, but also by their relative distances 
from the power-law fitting line which indicates the expected level of corruption. For example, Bulgaria and Slovenia 
differ significantly in their wealth (Slovenia has $\approx5$ times higher {\it gdp}), but both countries are at equal distances above the fitting
line, indicating (i) that both countries are less corrupt than the corruption level expected for their corresponding wealth and (ii) that the relative 
level of corruption of Slovenia within the group of countries falling in the same {\it gdp} bracket as Slovenia is similar to the relative corruption level of 
Bulgaria within the group of countries falling in the same {\it gdp} bracket as Bulgaria [Fig. 3].

%Depending whether a specific country is below (or above) the power law in Fig.~\ref{fig.1} we obtain for all countries, one can assess if this 
%country is more (or less) corrupt relative to the expected average level of corruption which corresponds to the country
%wealth measured by the {\it gdp}.

To test how robust is the power-law dependence between corruption and country wealth, we analyze groups containing different numbers of countries, and we find 
that Eq. (1) holds, with similar values of $\mu$ [Fig.~\ref{fig.1}(a-e)]. 
Averaging the power-law exponent $\mu$ for different years and for different number of countries we find $\overline\mu\approx$0.27$\pm\Delta$, 
            where $\Delta$=0.02 is the standard deviation. For the CPI and {\it gdp} data we find an average correlation coefficient of 0.86. 
            We also note that the inverse relation of {\it gdp} as a function of CPI is characterized by an exponent $\hat{\mu}$ which is not equal to 1/$\mu$ as one
            might expect, since the correlation coefficient of the data fit is less than 1. 
Next, we analyze data comprising the same set of countries for different years [Fig. 2], and we find 
that the power-law
dependence of Eq. (1) remains stable in time over periods shorter than a decade, with similar and slightly decreasing values for $\mu$ [Fig.~\ref{fig.1} and Fig. 2].
Similar results we obtain also for the period 1996-2000 (not shown in the figures as available data cover much smaller number of countries for that period).

Given the facts that (i) the number of countries we analyze changes from 90 to 153, 
and (ii) that the time horizon of 5-6 years we consider could be sufficient for significant changes in both corruption level and wealth (e.g., the case
of Eastern European countries), our finding of a power-law relationship in Eq. (1) is consistent with a 
universal dependence between {\it gdp} and CPI across diverse countries.
We note that the power-law relation in Eq. (1) holds when {\it gdp} is calculated both as current prices in US dollars [Fig. 1 and Fig. 2], as well
as the value based on purchasing power parity [Fig. 4].
Further, Eq. (1) implies that lowering the corruption level of a country would 
lead to an increase in its {\it gdp} and vice versa---e.g., for a country with {\it gdp} $\approx\$4000$
 an increase in CPI of 0.25 units would lead to increase in the {\it gdp} of approximately $\$700$ [Fig.~\ref{fig.1} and Fig. 2]. 
\begin{figure}
 \centering
   \rotatebox{-90}{\resizebox{0.85\columnwidth}{!}{\includegraphics{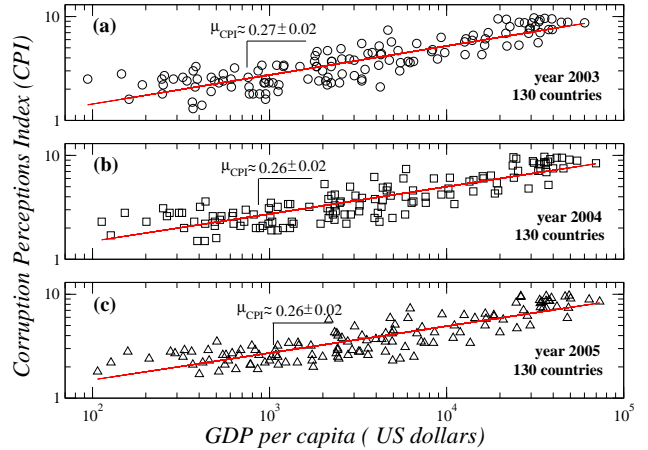}}}
\caption{Log-log plots of the corruption perceptions index (CPI) versus GDP per capita ({\it gdp}) for a subset of 130 countries over the period 2003-2005.
         The same set of countries is presented in each plot, indicating that the power-law exponent $\mu$ characterizing the 
         relation between CPI and {\it gdp} remains relatively stable over the considered period. As rich countries have a relatively higher
         {\it gdp} growth rate compared to poor countries (see Fig. 8 below), and because CPI is defined in a bounded interval,
         we expect the value of the exponent $\mu$ to decrease slightly with time when considering time horizons larger than a decade.
         }
\label{fig.1_130}
 \end{figure}
\newpage

\begin{figure}
 \centering
\rotatebox{-90}{\resizebox{0.8\columnwidth}{!}{\includegraphics{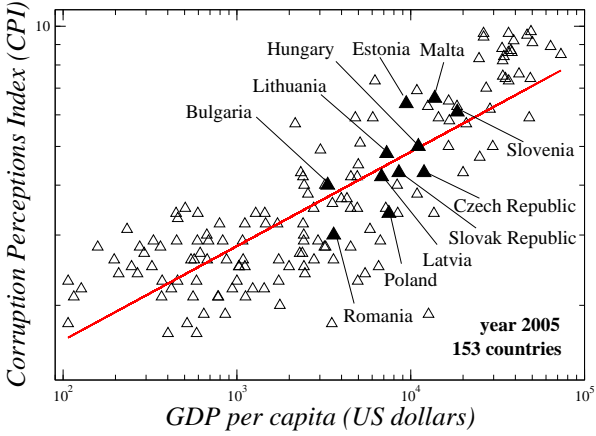}}}
\caption{ Same as panel (e) in Fig. 1 except we now identify
           by filled symbols the subset of the 153 countries, which are recently-accepted members of the
           European Union and candidates. 
            Although this subset varies greatly in wealth and corruption level, data also follow a similar scale-invariant behavior.
         } 
\label{fig.1g}
\end{figure}

\begin{figure}
 \centering
\rotatebox{-90}{\resizebox{0.85\columnwidth}{!}{\includegraphics{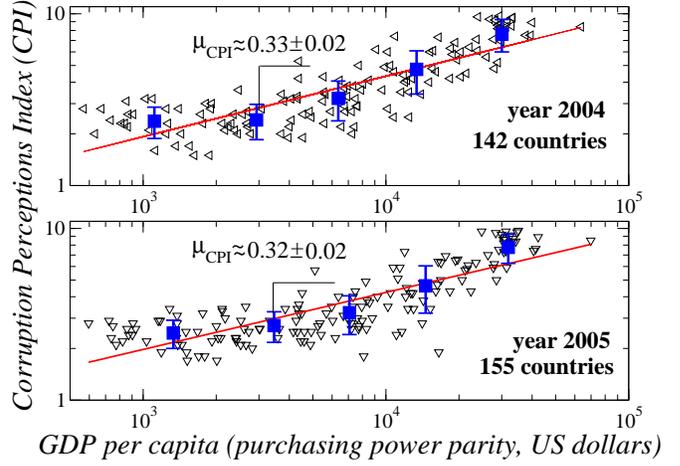}}}
\caption{ Log-log plots of the corruption perceptions index (CPI) versus GDP per capita ({\it gdp}) for the same years as shown in panels (d) and (e) in Fig. 1,
            indicating a power-law functional 
            dependence similar to Fig. 1. Data on {\it gdp} are obtained based on purchasing power parity in U.S dollars \cite{IMFGDP}. 
             A low value of CPI corresponds to a high level of corruption \cite{CPI}.             
              The power-law relation between CPI and {\it gdp} remains stable also for constant prices across different years and different number of countries,
            and is characterized by a similar value of the exponent $\mu$ as for current prices. We note that the slightly higher value
            of $\mu_{CPI}$ observed here compared with Fig. 1 and Fig. 2 is due to the slight reduction in the different between
           wealthy and poor countries when {\it gdp} is measured based on purchasing power parity. 
            The group average and standard deviation of the CPI for five subgroups of countries for both years are shown with filled squares. 
           The power-law fit across all countries indicates the expected level of corruption for a given range of country wealth.
        } 
\label{fig.1h}
\end{figure}

To confirm that our findings do not depend on the specific 
choice of the measure of corruption, we repeat our analysis for a different index, the CCI \cite{KKM,CC}. As the CCI is defined in the 
interval [--2.5, 2.5] we use a linear transformation to obtain the {\it adjusted} CCI,
 $CCI_{ad}~\equiv~2\times(CCI+2.5)$, so that both $CCI_{ad}$ and CPI are defined in the same interval from 
0 to 10. We find that $CCI_{ad}$ also exhibits a power-law behavior as a function of {\it gdp} with a similar value of the power-law exponent $\mu$ as 
obtained for CPI [Fig. 1(f)]. So, the specific interval in which the corruption index is defined does not affect the nature of our findings.

We note that there is no artificially imposed scale on the values of the CPI or CCI index for different countries. While the upper and lower bounds for 
the CPI or CCI index are 
indeed pre-determined, the intrinsic relative relation 
between the index values for different countries is inherent to the data. There is no logarithmic
 scale artificially imposed on the index values of each country (see details 
on the CPI and CCI methodology in \cite{CPI,CC,method}).
The fact that we obtain practically identical results (power-law dependence with similar values of the exponent $\mu$) 
 for two independent indices CPI and CCI, which are provided by different institutions and are calculated using different methodologies, 
indicates that the quantitative relation of Eq. (1) is not an artifact of subjective 
evaluation of corruption. In summary, our empirical results indicate that the power-law relation between 
corruption and {\it gdp} across countries does not depend on the 
specific subset of chosen countries (provided they span a broad range of {\it gdp}), does not depend on the specific measure of corruption (CPI and CCI), and
does not change significantly over time horizons shorter than a decade.

\subsection{Corruption Level and Country Wealth Rank Curves.} 
\begin{figure*}
  \centering
   \rotatebox{-90}{\resizebox{1.5\columnwidth}{!}{\includegraphics{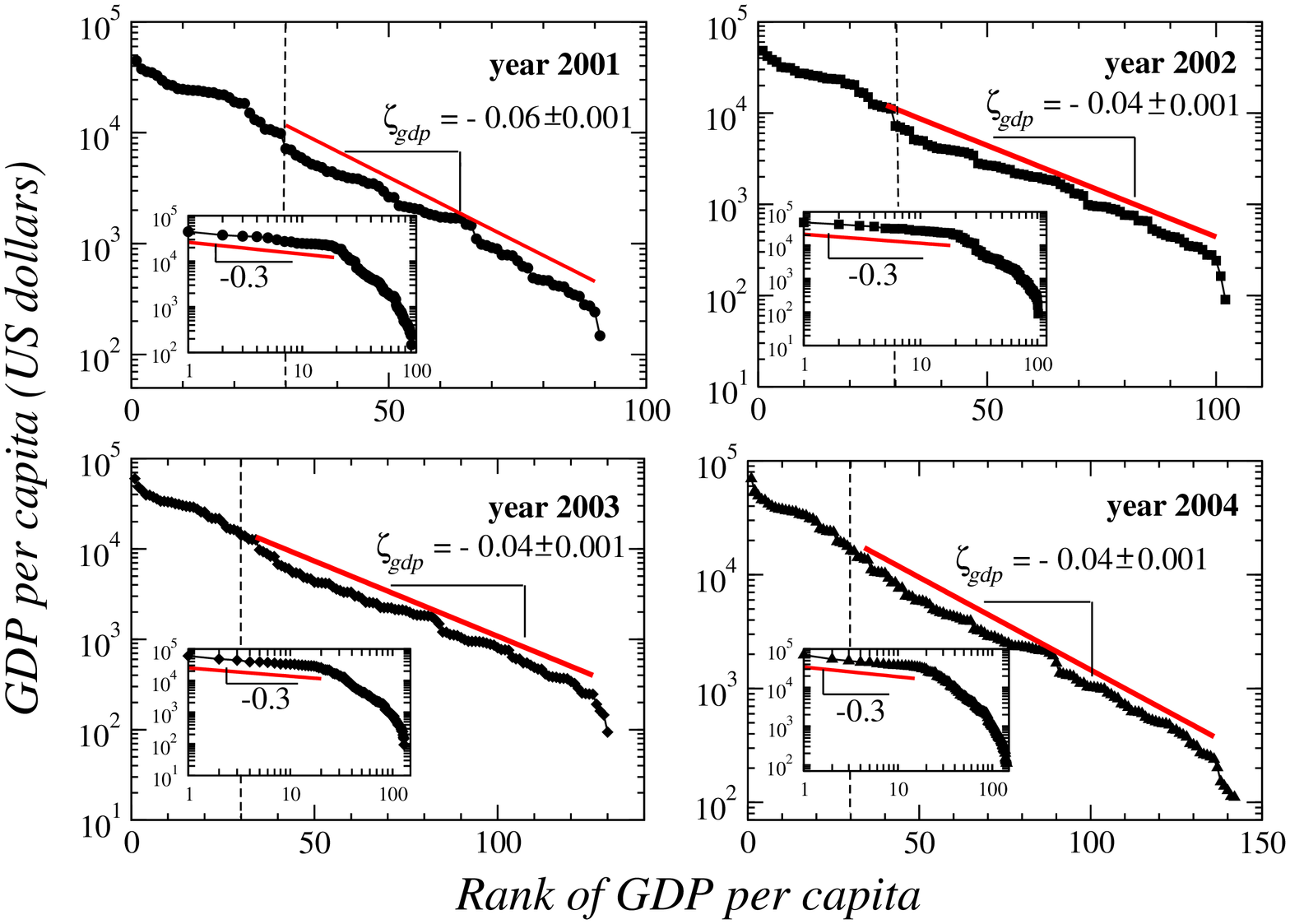}}}
  %\vspace{-0.5cm}
\centering
  %\vspace{-0.5cm}
   \rotatebox{-90}{\resizebox{0.7\columnwidth}{!}{\includegraphics{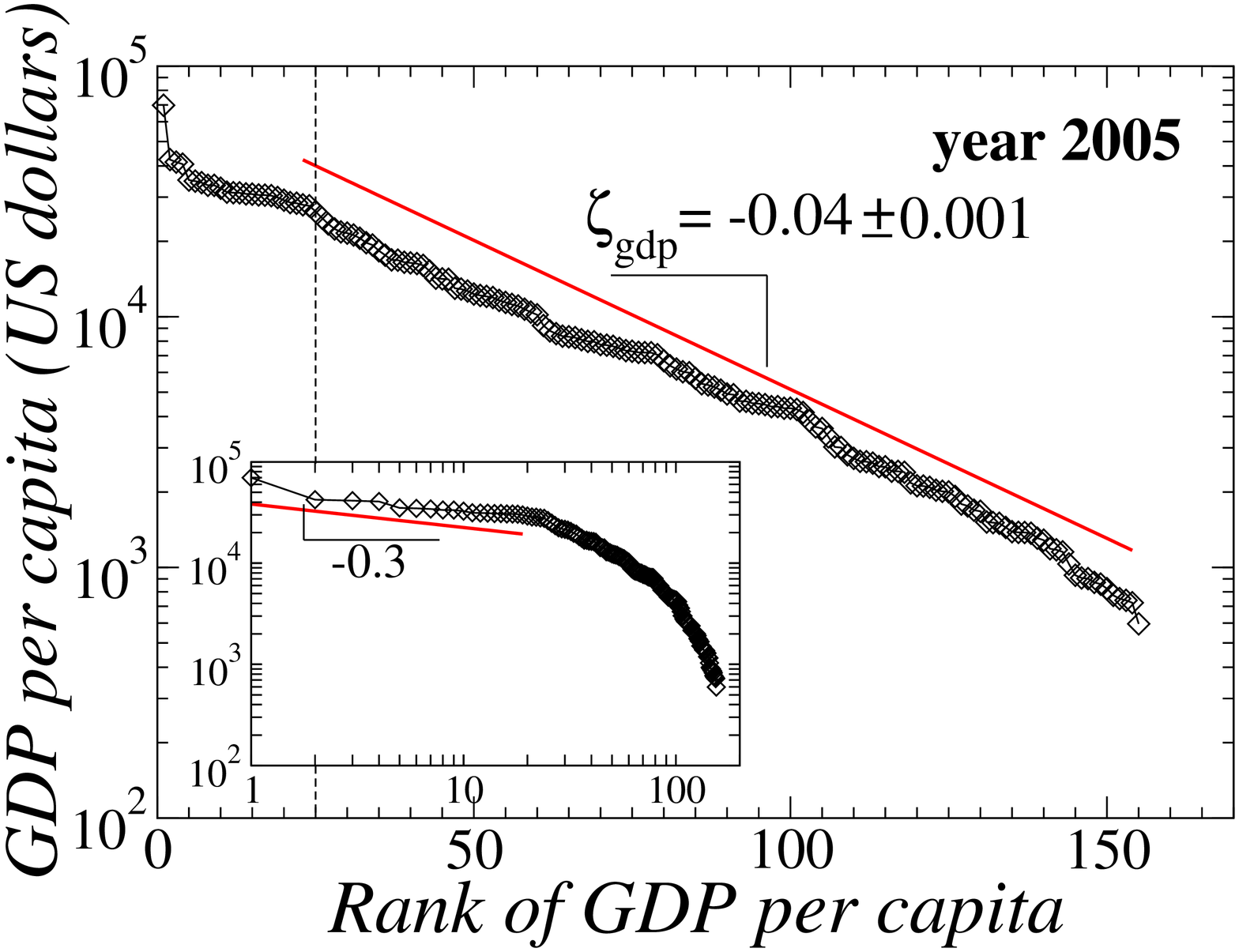}}}
%\vspace{-1cm}
  % \rotatebox{-90}{\resizebox{0.8\columnwidth}{!}{\includegraphics{Fig2bs.ps}}}
  % \rotatebox{-90}{\resizebox{0.5\columnwidth}{!}{\includegraphics{rank_2005cpi.ps}}}
    \caption{ Zipf plots ranking in decreasing order the GDP per capita ({\it gdp}) for the same 
            groups of countries and for the same years as shown in 
            Fig. 1. Data on GDP 
            per capita are obtained from the International Monetary Fund as current prices in U.S. dollars \cite{IMFGDP}.
            Fitting lines indicate exponential behavior for the GDP per capita for countries below rank 30 (vertical dashed line, shown also in Fig. 1), 
            characterized by the exponential decay constant $\zeta_{\it gdp}$. 
           % The ratio $\zeta_{CPI}$/$\zeta_{\it gdp}$ consistently reproduces the 
           % value of the power-law exponent $\mu$ in Fig.1(a) - Fig.1(d) for each corresponding year and each group of 
           % countries. This indicates that a necessary condition for the power-law relation between CPI and GDP per capita is that the GDP per capita rank order
           % of countries is similar to the rank order based on CPI. Log-log plots of the ranking curves (shown in the insets) indicate a crossover from an exponential
           % to a power-law behaviour for the top 30 wealthiest countries as well as for the top 30 least corrupt countries. 
           Log-log plots of the ranking curves (shown in the insets) indicate a crossover from an exponential
            to a power-law behaviour for the top 30 wealthiest countries.
            We note that the top 30 wealthiest 
            countries cluster above the fitting curves in Fig. 1, Fig. 2 and Fig. 4.
            }
 \label{fig.2}
 \end{figure*}

\begin{figure*}
 \centering
 \vspace{-1.5cm}
\rotatebox{-90}{\resizebox{1.4\columnwidth}{!}{\includegraphics{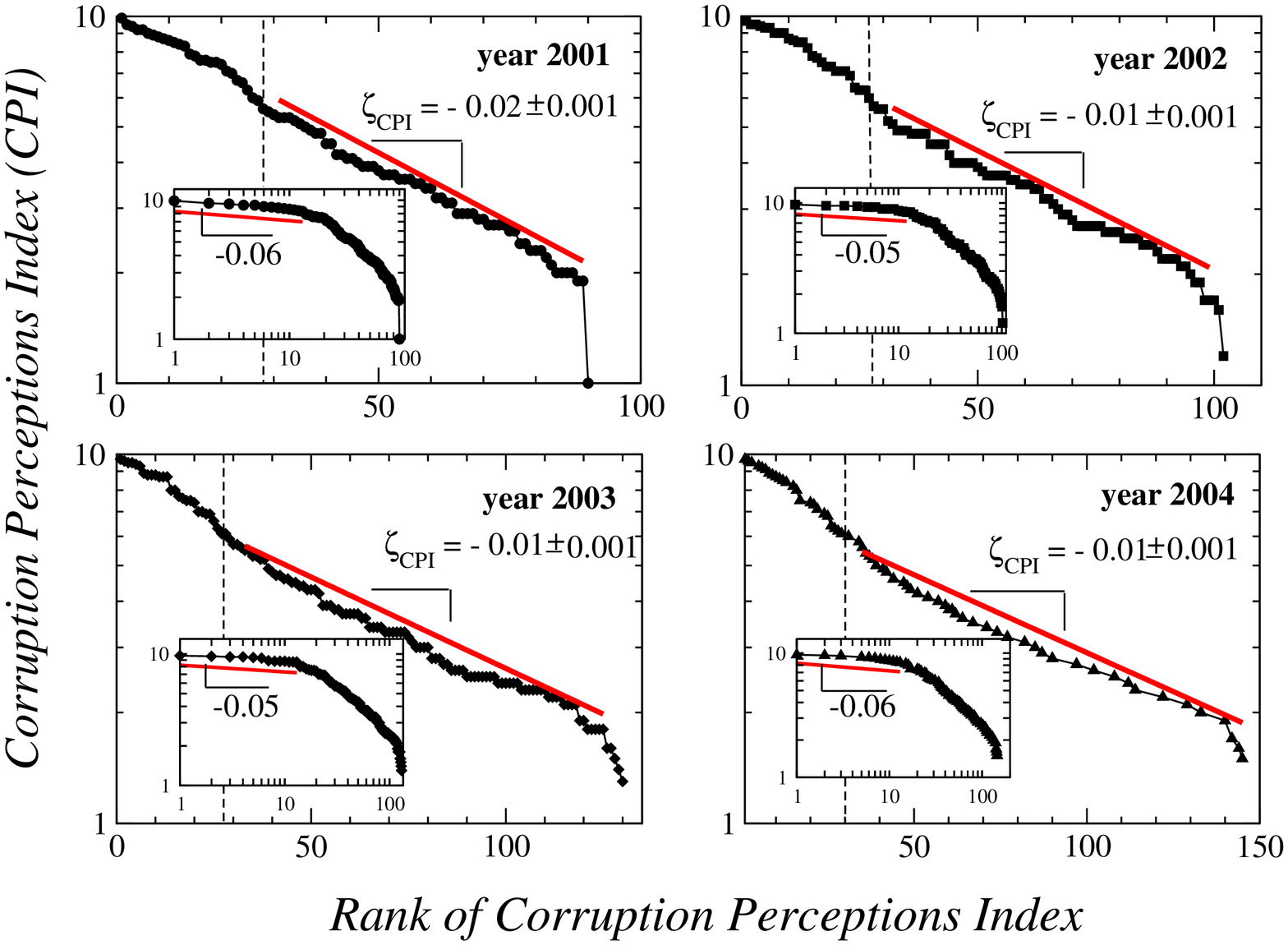}}}
 \centering
   \rotatebox{-90}{\resizebox{0.6\columnwidth}{!}{\includegraphics{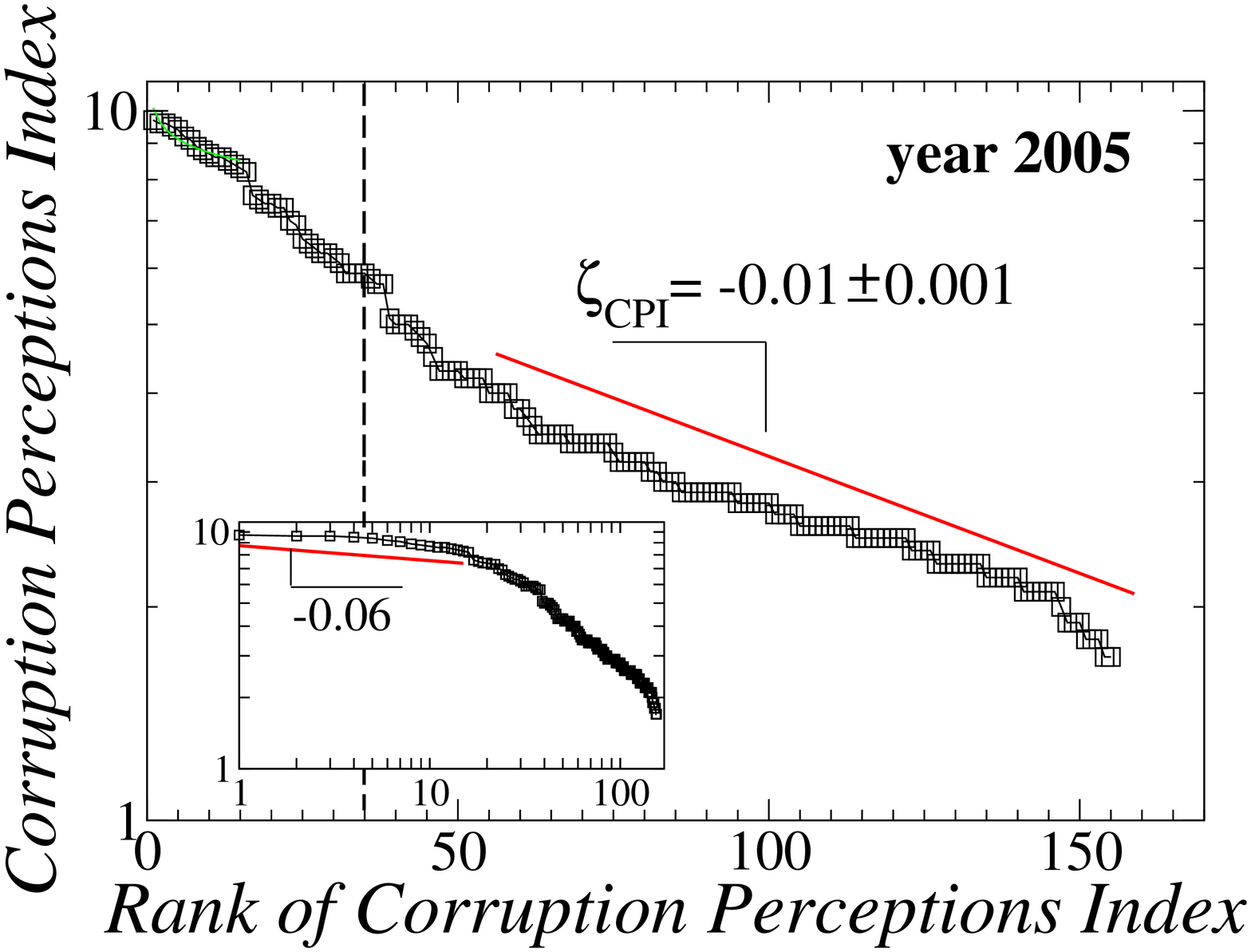}}}
   \caption{ Zipf plots ranking in decreasing order the CPI for the same 
            groups of countries and for the same years as shown in 
            Fig. 1 and Fig. 5.
            Fitting lines indicate exponential behavior for the CPI for countries below rank 30 (vertical dashed line, shown also in Fig. 1 and Fig. 5), 
            characterized by the exponential decay constant $\zeta_{CPI}$. 
            The ratio $\zeta_{CPI}$/$\zeta_{\it gdp}$ consistently reproduces the 
            value of the power-law exponent $\mu$ in Fig. 1(a) - Fig. 1(d) for each corresponding year and each group of 
            countries. This indicates that a necessary condition for the power-law relation between CPI and GDP per capita is that the GDP per capita rank order
            of countries is similar to the rank order based on CPI. Log-log plots of the ranking curves (shown in the insets) indicate a crossover from an exponential
            to a power-law behaviour for the top 30 least corrupt countries, similar to the cross over behaviour observed for {\it gdp} in Fig. 5.
          }
\label{fig.2b}
\end{figure*}

We next rank countries by their {\it gdp} and 
by their CPI. We find that {\it gdp} versus 
rank exhibits an exponential behavior for countries with rank larger than 30, and a pronounced 
crossover to a power-law behavior for the wealthiest 30
countries [Fig.~\ref{fig.2}]. We further find that the shape of {\it gdp} versus rank curve remains unchanged for different years, and that increasing the number of 
countries we consider 
only extends the range of the exponential tail. Our findings for the shape of the {\it gdp} versus rank curve differ from earlier reports
 \cite{zipf1,zipf2}. We find that the CPI versus rank curve exhibits a behavior similarly to that of the {\it gdp} versus rank curve, with 
a crossover from a power law to an exponential tail for countries with rank larger than 30. 
The shape of the CPI versus rank curve also remains unchanged when we repeat the analysis for different years [Fig.~\ref{fig.2b}]. 
 We find that the ranking of countries based on 
{\it gdp} practically matches the ranking based on the CPI index. 
This is evidence of a strong and positive correlation between the ranking of wealth and the ranking of corruption.
 Since the {\it gdp} rank is an unambiguous result of an {\it objective} quantitative measure, the evidence of a strong correlation of the CPI 
rank with the {\it gdp} rank we observe in Fig.~\ref{fig.2} and Fig.~\ref{fig.2b} 
 indicates that the CPI values are not {\it subjective}, and that our finding of a power-law
relation between CPI and {\it gdp} in Fig. 1 and Fig. 2 is not an artifact of an arbitrary scale imposed on the CPI or on the CCI.
Further, we compare the values of the decay parameters $\zeta_{CPI}$ and $\zeta_{\it gdp}$ characterizing the 
exponential behavior of the CPI and {\it gdp} rank curves,
\begin{equation}
      CPI\sim \exp(\zeta_{CPI}\cdot R_{CPI}),
\end{equation}
and
\begin{equation}
      {\it gdp}\sim \exp(\zeta_{\it gdp}\cdot R_{\it gdp}),
\end{equation}
where $R_{CPI}$ and $R_{\it gdp}$ index the rank order of CPI and {\it gdp} respectively.

We find that for each year the ratio $\zeta_{CPI}$/$\zeta_{\it gdp}$ reproduces the value of the power-law exponent $\mu$ defined in Eq. (1) 
for the same year ---
an insightful result since it would hold only when $R_{CPI}$ is similar to $R_{\it gdp}$.
Indeed, only when $R_{CPI}\approx R_{\it gdp}$ we obtain from Eq. (2) and Eq. (3) the relation between log(CPI) and log({\it gdp}),
\begin{equation}
     log(CPI)\approx (\zeta_{CPI}/\zeta_{\it gdp})\cdot log({\it gdp}).
\end{equation}
Combining Eq. (1) and Eq. (4), we see that
\begin{equation}
 \mu=\zeta_{CPI}/\zeta_{\it gdp}.
\end{equation}
%linking the ratio of the parameters $\zeta_{CPI}$ and $\zeta_{\it gdp}$ with the power-law  $\mu$ in Eq. (1).
Thus, for each year the power-law dependence between CPI and {\it gdp} in Eq. (1) is
 directly related to the exponential behavior of the CPI and {\it gdp} versus rank [Eq. (2) and Eq. (3)]. We note that
this relation does not hold for the top 30 wealthiest countries, for which there is an enhanced economic interaction in a globalization 
sense, perhaps leading to similarities in development patterns and overall decrease in the {\it gdp} growth difference \cite{asloos05,asloos06}

%\begin{equation}
%      CPI\sim 10^{(\zeta_{CPI}\cdot R_{CPI})},
%\end{equation}
%\begin{equation}
%      {\it gdp}\sim 10^{(\zeta_{\it gdp}\cdot R_{\it gdp})},
%\end{equation}
%where $R_{CPI}$ and $R_{\it gdp}$ represent ranking of CPI and {\it gdp} respectively, and also we have:      
%\begin{equation}
%      R_{CPI}\sim R_{\it gdp},
%\end{equation}
%we can get the relation between log(CPI) and log({\it gdp}):
%\begin{equation}
%     log(CPI)\sim (\zeta_{CPI}/\zeta_{\it gdp})\cdot log({\it gdp}),
%\end{equation}
%linking with Eq.(1), which implies
%\begin{equation}
%     \mu=\zeta_{CPI}/\zeta_{\it gdp},
%\end{equation}

%\newpage

\subsection{Relation Between Corruption Level and Foreign Direct Investment.} 
\begin{figure*}
 \centering
  \rotatebox{-90}{\resizebox{1.3\columnwidth}{!}{\includegraphics{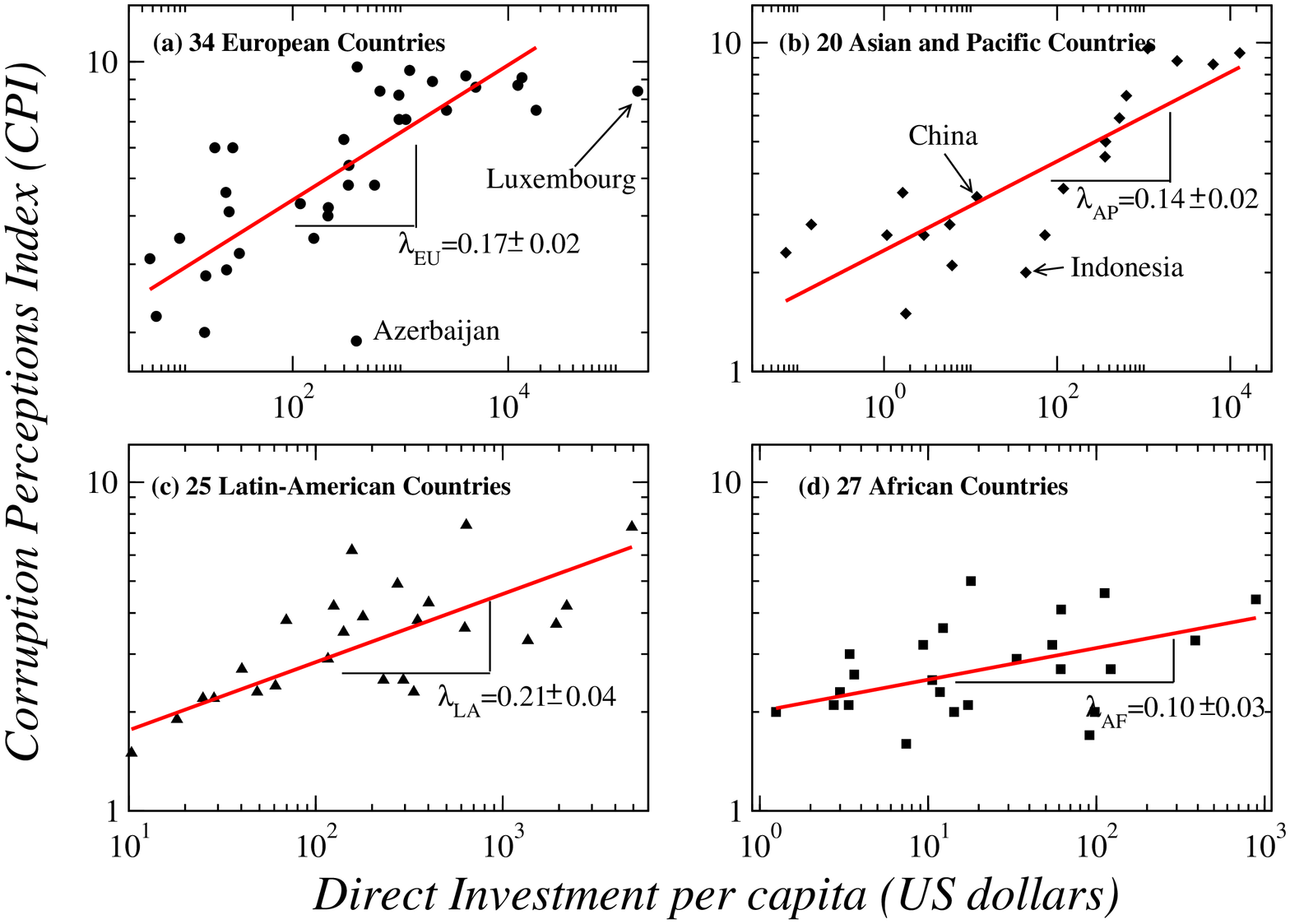}}}
 \caption{Log-log plots of the CPI versus the amount of direct investment on a historical-cost basis from the United States received by
   different countries for the year 2004 \cite{BEA}.
  We observe strong positive correlation between level of investment per capita and level of corruption --- countries 
  with high CPI receive also larger investment. Shown are (a) 34 
  European countries, (b) 20 Asian-Pacific countries, (c) 25 Latin-American countries and (d) 27 African countries. Note the striking difference between
  the typical values of direct investment per capita when comparing, say, European countries and African countries, with typical values of $CPI\approx5$ and
  $CPI\approx2.5$ respectively. The correlation coefficients of the fits in (a), (b), (c) and (d) are 0.74, 0.83, 0.69 and 0.37 respectively. Note that although
  China receives a huge net inflow of U.S. investment each year, the per capita investment from the U.S. is not very high, and is quite similar to the U.S.
  per capita investment for countries with a CPI value similar to that of China.
   }
\label{fig.3} 
\end{figure*}

We next investigate 
how the corruption level relates to foreign direct investment. We consider the amount of inward investments 
received by different countries from 
the United States (U.S.). Investments originating from the U.S. are sensitive to corruption, since U.S. legislation 
holds American investors in other countries liable for corruption 
practices \cite{FCPA}. We find a strong dependence of the amount of U.S. direct investments in a given 
country on the corruption level in that country [Fig.~\ref{fig.3}].
 Specifically, we find that the functional dependence between U.S. direct investments per capita, {\it I}, and the corruption levels
across countries exhibits 
scale-invariant behavior characterized by a power law ranging over at least a factor of $10^{3}$ [Fig.~\ref{fig.3}]
\begin{equation}
     CPI\sim {\it I}^{\lambda}.
\end{equation}
We find that less corrupt countries 
have received more U.S. investment per capita, and that Eq. (6) also holds for different years. 
%Further, we find that the scaling exponent $\lambda$ significantly changes for groups of countries with different average CPI.
In particular, we find that groups of countries from different continents, which differ in both {\it gdp} and average CPI, are characterized by different 
values of $\lambda$ [Fig.~\ref{fig.3}]. We obtain similar results when repeating our analysis 
for the CCI, suggesting that the power-law relation in Eq. (6) 
between corruption level and foreign direct investment per capita does not depend on the specific measure of corruption used.
We also note that the 1977 Foreign Corrupt Practices Act \cite{FCPA} only precludes American firms from entering corruption deals, but does not dictate
in which country and how much money the American firms should invest. Therefore, the statistical regularities 
we find in Fig.~\ref{fig.3} cannot arise 
from legislatory measures against foreign corruption.

\subsection{Relation Between Corruption Level and Growth Rate.} Finally, we investigate whether 
there is a relation between corruption level and long-term growth rate. Since the CPI reflects the 
quality of governing and administration in a given country, which traditionally requires considerable time to change, we hypothesize that there may be  
relation between the current corruption level 
of a country and 
its growth rate over a wide range of time horizons. To test this hypothesis we estimate the long-term growth rate for each country as the slope of the least 
square fit to the plot of log({\it gdp}) 
versus year over the past several decades,
where the {\it gdp} is taken as constant prices in national currency [Fig. 8]. 
 We divide all countries into four groups according to the World Bank classification based on {\it gdp} \cite{CLASS}. 
We find a 
strong positive dependence between country group average of CPI and the group average long-term growth rate, showing that less corrupt countries  
exhibit significant economic growth 
while more corrupt countries display insignificant growth rates (or even display negative growth rates) [Fig.~\ref{fig.4}]. Repeating our analysis for 
different time horizons (1990-2005;
 1980-2005) we find similar relations between the CPI and the long-term growth, indicating a link between corruption and economic growth.

\begin{figure}[h!]
 \centering
  \rotatebox{-90}{\resizebox{0.7\columnwidth}{!}{\includegraphics{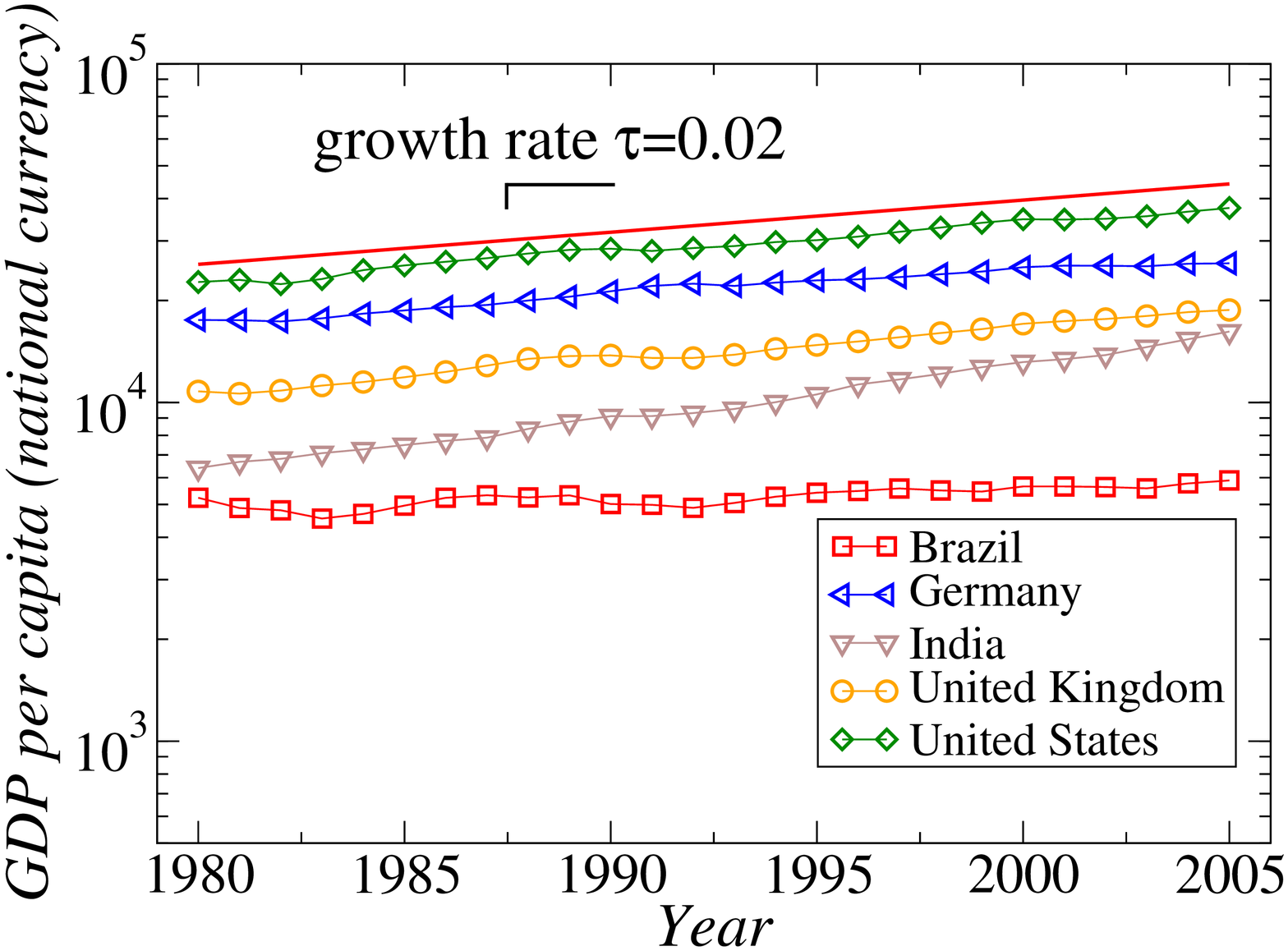}}}
\caption{Long-term growth rate of the GDP per capita ({\it gdp}) measures as constant prices in national currency \cite{IMFGDP} over the period 1980 to 2005.
 Seperate curves represent countries of different wealth and corruption level from different continents. All countries exhibit exponential growth characterized 
by average long-term growth rate $\tau$, estimated for each country as the slope of the least square fit to the plot of log({\it gdp}) versus year over the
period 1980 to 2005. The fitting line indicates the long-term growth rate $\tau$ of United States over the period 1980 to 2005.
}
\label{fig.4a}
\end{figure}

\begin{figure}[h!]
\centering
  \rotatebox{-90}{\resizebox{0.7\columnwidth}{!}{\includegraphics{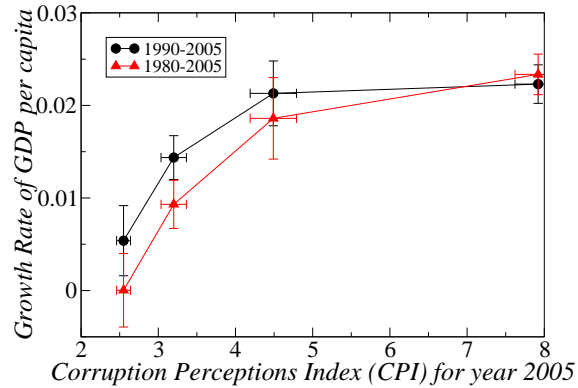}}}
 \caption{ Relation between the CPI for the year 2005 \cite{CPI} and the long-term growth rate of GDP per capita ({\it gdp}) as constant 
    prices in national currency \cite{IMFGDP}. Each curve represents 120 countries
    divided into 4 groups based on their level of wealth 
    according to the World Bank's classification \cite{CLASS}. Excluded are countries with population less than two million and 
    countries for which the GDP per capita records extend back fewer than 15 years starting from 2005. The long-term growth rate is 
    estimated over the periods 1980-2005 and 1990-2005.
    Symbols represent the group 
    average value of CPI and long-term growth rate of GDP per capita. The error bars represent the group standard error. 
    The plot indicates that the corruption level of a country at present is strongly related to the past long-term growth rate of the 
    GDP per capita. Countries which are presently more
    corrupt exhibit on average negligible or even negative growth rates. In contrast, less corrupt countries exhibit
    higher growth rates. This strong correlation between corruption level and growth rate of GDP per capita remains valid for a broad range of past
    time horizons.
    %Note that the upper-middle wealth countries 
    %with average CPI$\approx$4.5 show
    %the most significant change in GDP per capita growth rate, compared to the other three groups.
     }
\label{fig.4}
\end{figure}

In summary, the functional relations we report 
here can have implications when determining the relative level of corruption between countries, and for 
quantifying the impact of corruption when planning foreign investments and economic growth. These quantitative relations may further facilitate current studies 
on spread of corruption across social networks \cite{blanchard}, the emergence of endogenous transitions from one level of corruption to another through cascades of 
agent-based micro-level interactions \cite{ross,Situngkir}, as well as when considering corruption in the context of certain cultural norms \cite{fisman}.

\vspace*{1.0cm}

ACKNOWLEDGMENTS: We thank F. Liljeros for valuable suggestions and discussions, and we thank D. Schmitt and F. Pammolli for helpful comments.
We also thank Merck Foundation, NSF, and NIH for financial support.

\vspace*{-0.3cm}

\end{document}